\documentstyle[multicol,aps,prl,epsfig]{revtex}
\begin{document}
\draft
\title{Giant Magnetoresistance at the Interface of Iron Thin Films}
\wideabs{
\author{J. Balogh, L.F. Kiss}
\address{Research Institute for Solid State Physics and Optics,\\ 
H-1525 Budapest, P.O.B. 49, Hungary}
\author{A. Halbritter, I. K\'{e}zsm\'{a}rki and G. Mih\'{a}ly}
\address{Department of Physics, Institute of Physics \\
Budapest University of Technology and Economics, H-1111 Budapest, Budafoki \'{u}t 8, Hungary}
\date{\today}
\maketitle

\begin{abstract}
Ag/Fe/Ag and Cr/Fe/Cr trilayers with a single $25\ nm$ thick ferromagnetic layer 
exhibit giant magnetoresistance (GMR) type behavior. The resistance decreases for 
parallel and transversal magnetic field alignements with a Langevin-type magnetic 
field dependence up to $B=12$\ T. The phenomenon is explained by a granular 
interface structure. Results on Fe/Ag multilayers are also interpreted in terms 
of a granular interface magnetoresistance.
\end{abstract}

}

Magnetoresistance arising from a nanoscale magnetic structure was first
discovered in antiferromagnetically coupled multilayers prepared by
molecular beam epitaxy (MBE) \cite{B1,B2}. It is generally referred to as giant
magnetoresistance (GMR) due to the large resistance change as compared to
the anisotropic magnetoresistance (AMR) of bulk ferromagnetic materials. The
large effect gives great potential for technological applications, however,
as concerning the underlying physics the
absence of anisotropy in case of the GMR effect \cite{B1} is more important. 
This means that
$[R(H)-R(0)]/R(0)$ is negative irrespectively of the direction of the
measuring current and the applied magnetic field when both lie in the sample
plane. Since the first reports on GMR it turned out to be a more complex
phenomenon, as it was observed in polycrystalline multilayers \cite{P1}, granular
materials \cite{X1,B3}, magnetic domain walls \cite{G1} or supersaturated alloys 
\cite{O1}, as well. The criteria for observing GMR can be put more generally: the
characteristic length scales of the magnetic inhomogenities should be in the
order of the electronic mean free path. However, the respective role of
different bulk and interface scattering processes \cite{Z1} is not yet clear even
in case of multilayers. Many recent experimental \cite{S1,V1} and theoretical
\cite{Z1,T1} works addressed this question, but the role of the interface was
examined only in conjunction with the multilayer structure. 

In this letter
we show that by studying the magnetoresistance of trilayers containing only one 
magnetic layer it is possible to separate the contribution of a
single interface. The interface magnetoresistance has GMR characteristics and 
is attributed 
to a granular interface structure. The existance of a granular interface 
magnetoresistance raises the question how this term is 
related to the magnetoresistance arising from interlayer coupling in case of multilayers.   
Our study demonstrates that the granular interface contribution is dominant in 
Fe/Ag multilayers. By investigating Cr/Fe/Cr trilayers it is also shown that the 
granular interface magnetoresistance is not restricted to immiscible elements. 

The trilayer and
multilayer samples were prepared on Si single crystal substrate at room
temperature by vacuum evaporation in a base pressure of $10^{-7}$\ Pa. The layer
thickness was controlled by a quartz oscillator during sample deposition.
The magnetoresistance was measured by four contact method on $2$\ mm thick and $10$\ mm
long samples with current in the plane geometry. The magnetic field was
applied in three geometries: i., in the sample plane parallel to the current
ii., in the sample plane perpendicular to the current and iii.,
perpendicular to the sample plane. Magnetoresistance measured in the above
geometries are usually called parallel ($R_{\parallel}$), transversal ($R_{\perp}$) 
and perpendicular ($R_{perp}$),
respectively. The layer thickness range of the trilayers ($8nm$ for Ag and
Cr and $25nm$ for Fe) was small enough that the interface
magnetoresistance was not shunted by the resistance of the layers, however
it was thick enough that the Fe layer shows magnetic properties (saturation
magnetization, Curie temperature, demagnetization field) similar to bulk
layers. 

\begin{figure}
\noindent
\centerline{\includegraphics[width=0.78\linewidth]{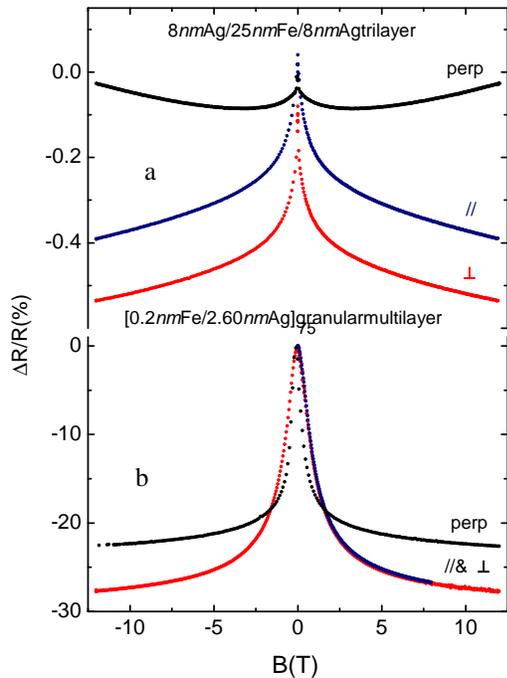}}
\caption{\it Magnetoresistance measured at $T=4.2$\ K with magnetic field
aligned parallel ($\parallel$), transversal ($\perp$) and perpendicular ($perp$)
to the measuring current for a $8nm\ $Ag/$25nm\ $Fe/$8nm\ $Ag trilayer
(a) and a granular sample prepared as a discontinuous multilayer of 
[$0.2nm$\ Fe\ +\ $2.6nm$\ Ag]$_{75}$ nominal sequence (b).}
\label{Figure 1}
\end{figure}

Magnetoresistance~curve~of~the~as~deposited $8nm\ $Ag/$25nm\ $Fe/$8nm\ $Ag 
trilayer measured in parallel, transversal and perpendicular geometries
up to $B=12$\ T 
magnetic field at $T=4.2$\ K is shown in Fig.\ \ref{Figure 1}a. $R_{\parallel}$ 
and $R_{\perp}$ show similar 
magnetic field dependence in the high
field region, but the two curves are shifted relative to each other because
of a small AMR below $0.2$\ T. The lower value of $R_{\perp}$ is consistent with
thin film measurements on Fe \cite{G2} indicating a crossover in the signe 
of the parallel and the transversal magnetoresistance around this layer thickness. 
However, the high
field behavior is rather unusual. The equal decrease of the parallel and the
transversal magnetoresistance and the absence of saturation up to $12$\ T
magnetic field have not yet been observed on a single ferromagnetic Fe layer.
On the other hand the cusp like shape of the magnetoresistance curves and the
extremely high saturation field is typical of granular systems \cite{B3,X1}.

Magnetoresistance of a Fe-Ag sample which has a 
[$0.2nm$\ Fe\ +\ $2.6nm$\ Ag]$_{75}$  nominal  
multilayer structure is shown in Fig.\ \ref{Figure 1}b. The thin Fe layers 
are not continuous in this sample and this speciman shows characteristics 
of a granular system, 
e.g. it is superparamagnetic with a blocking temperature 
around $40\ K$ \cite{Kiss}. The magnetic field dependence of $R_{\parallel}$ and $R_{\perp}$ 
also shows the characteristic 
features observed on granular samples prepared by co-deposition of the constituents 
\cite{XXX}. A distinct feature of our granular sample prepared by sequential 
deposition is the anisotropy observed when the magnetic field is perpendicular 
to the sample plane. In co-deposited granular materials \cite{M1} there are 
only minor differences between $R_{\parallel}$, $R_{\perp}$ and $R_{perp}$. 
The layered growth seems to strongly affect the sample morphology 
and probably the shape of the granules. 
Note that $R_{perp}$  of the trilayer sample also shows distinct behavior
(see Fig.\ \ref{Figure 1}a). 

In granular materials the GMR phenomenon is attributed to spin dependent
scattering on single domain ferromagnetic particles and is shown to scale
with the reduced magnetization \cite{X1} as: 

\begin{equation}
\frac{R(H)-R(0)}{R(0)} = -A\left( {\frac{M} {M_s}} \right) ^2 ,
\label{eq:1}
\end{equation}

where $M$ is the global magnetization of the sample and $M_s$ is the saturation
magnetization. The prefactor $A$ depends on the number and the size of the
single domain particles. According to the classical theory of
superparamagnetism the reduced magnetization can be described by the
Langevin function, therefore 

\begin{equation}
\frac{R(H)-R(0)}{R(0)} =  -A{L^{2}}(mH/kT),
\label{eq:2}
\end{equation}

where
$L(x)=cth(x)-1/x$ and $m$ is the magnetic moment of the superparamagnetic
particles. Equation (\ref{eq:2}) was extended in order to be applicable for the trilayer 
and the multilayers, as well:
\begin{equation} 
\frac{R(H)-R(0)}{R(0)} =  -A_1{L}^2(mH/kT)-A_2H^2 + A_3 .
\label{eq:3}
\end{equation}
To account for 
possible scattering on single Fe impurities in the nonmagnetic matrix 
a term proportional to $H^2$ was included \cite{Majom}. The constant $A_3$  
describes the shift in the high field magnetoresistance due to the 
AMR contribution of ferromagnetic particles.  
All the experimental curves could be satisfactorily fitted with Eq.\ (\ref{eq:3}) 
in the $B>0.25$\ T range.

\begin{figure}
\noindent
\centerline{\includegraphics[width=0.78\linewidth]{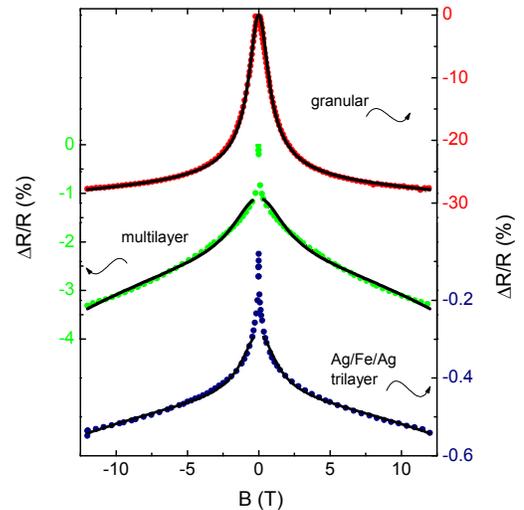}}
\caption{\it Magnetoresistance curves measured with transversal magnetic field
alignment at $4.2$\ K and fitted according to Eq.\ (\ref{eq:3}) in the $\left| B\right| >0.25$\ T
magnetic field range. Upper panel is for a granular sample prepared as a
discontinous multilayer of [$0.2nm$\ Fe\ +\ $2.6nm$\ Ag]$_{75}$  nominal
sequence. Middle panel is for a multilayer of [$1.4nm$\ Fe\ +\ $2nm$\ Ag]$_{60}$ 
nominal sequence with continuous ferromagnetic layers. Lower panel
is for a $8nm\ $Ag/$25nm\ $Fe/$8nm\ $Ag trilayer. Since the measured
points and the fitted curves mostly overlap on this scale the measured
points are rarefied for clarity.}
\label{Figure 2}
\end{figure}

Figure \ref{Figure 2} shows the fit results for three representative samples of different morphology. 
The first one is a granular sample, the same as in Fig.\ \ref{Figure 1}b, prepared by sequential deposition. 
The second one is a multilayer with continuous ferromagnetic layers of 
[$1.4nm$\ Fe\ +\ $2nm$\ Ag]$_{60}$ nominal sequence. For this thickness range a RKKY-type 
magnetic coupling is expected between the Fe layers \cite{Y1}.  The last one is a trilayer sample, 
the same as in Fig.\ \ref{Figure 1}a, containing one magnetic layer.

For the above three samples $m=19$, $7$, 
and $11\  {\mu}_B$ cluster moments are obtained, respectively. 
The magnetic moment of Fe atoms belonging to
different size Fe clusters in the Ag matrix \cite{N1} is about $3\ {\mu}_B$. According to 
the above analysis the magnetoresistance is determined by clusters containig a few (3-7) 
Fe atoms. We found that for all the samples the 
Langevin term is the dominant one ($A_1= 5.3\times 10^{-1}$, $1.3\times10^{-1}$ 
and $4.4\times 10^{-2}$, respectively).  
The AMR shift is zero for the granular sample, however it is non negligible for the multilayer 
and the trilayer ($A_3= -0.01$ and $ –0.003$, respectively) due to the continous 
ferromagnetic layers.
The quadratice term is relatively small in the granular sample ($A_2= 5\times 10^{-5}$), however 
it is non-negligible as compared to $A_1$ for the multilayer and the trilayer 
($A_2= 6\times 10^{-5}$ and $5\times 10^{-6}$, respectively). The above $A_2$ amplitudes can be   
associated with Fe impurities in the nonmagnetic matrix in the order of a few hundred ppm 
\cite{Majom}.  

According to the above analysis the
unusual high field magnetoresistance of the trilayer sample is attributed to
mixing of the Fe and Ag atoms at the interface and the formation of a
granular-like interface alloy. Fe and Ag are immiscible at equilibrium but
the substantially smaller surface free energy of silver makes an Ag covered
surface energetically favourable. It has been shown that this acts as a
driving force for Ag diffusion through ultra thin Fe layers, either during
sample deposition \cite{B4} on substrates at or above room temperature or during
a heat treatment \cite{S2} at low temperature ($200-300^\circ C$). However, if
interface mixing can occur in case of inmiscible elements it is even more
likely for constituents with a limited solubility and the question can be
put forward, if an interface magnetoresistance is to be observed, as well.
To answer this question the Fe-Cr system was studied. The phase diagram of
the Fe-Cr system shows solubility above 1094~K in the entire concentration
range and at room temperature the solubility limit is a few at\%\ on each side.

\begin{figure}
\noindent
\centerline{\includegraphics[width=0.9\linewidth]{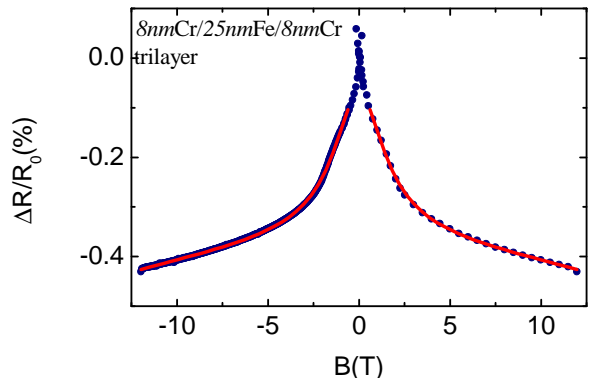}}
\caption{\it Transversal magnetoresistance of a $8nm$\ Cr/$25nm$\ Fe/$8nm$\
Cr trilayer measured at $4.2$\ K. The fitted curve corresponds to Eq.\ (\ref{eq:3}) in the $\left| B\right|
>0.25$\ T range. For  $B>0$\ T the measured points are rarefied for clarity.}
\label{Figure 3}
\end{figure}

Magnetoresistance measured on the $8nm$\ Cr/$25nm$\ Fe/$8nm$\ Cr trilayer 
is shown in Fig.\ \ref{Figure 3}.

The granular type magnetoresistance can also be observed 
and could also be fitted according to Eq.\ (\ref{eq:3}) ($m=9\  {\mu}_B$, $A_2=3\times 
10^{-6}$ and 
$A_3=-0.001$) as shown in Fig.\ \ref{Figure 3}. Non-equilibrium 
alloying at the interface can also be an adequate explanation, since similar
behavior was observed in supersaturated bulk Fe-Cr alloys \cite{O1} and in
co-sputtered alloy films \cite{S3}. 

In case of the relatively large layer
thickness of our Ag/Fe/Ag trilayer a demixing of the interface alloy can be
achieved by annealing without basically destroying the layered geometry.

\begin{figure}
\noindent
\centerline{\includegraphics[width=0.83\linewidth]{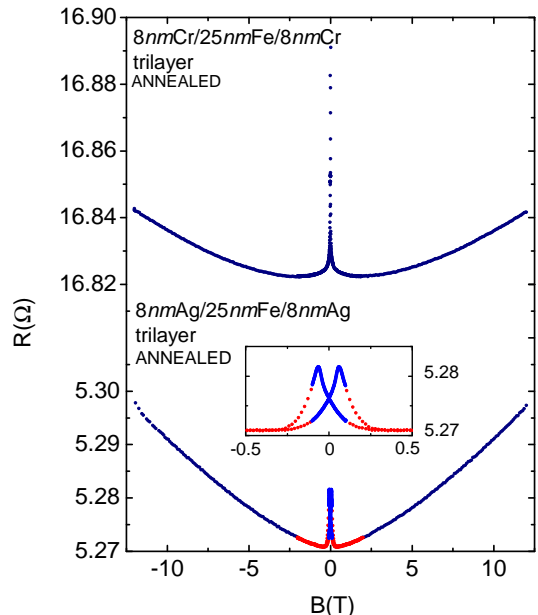}}
\caption{\it Magnetoresistance measured $T=4.2$\ K on the $8nm$\ Cr/$25nm$\ Fe/$8nm$\ Cr
and the $8nm\ $Ag/$25nm\ $Fe/$8nm\ $Ag trilayers after a heat
treatment in vacuum at $500^\circ C$. The insert shows the low field
behavior for the $8nm\ $Ag/$25nm\ $Fe/$8nm\ $Ag trilayer.}
\label{Figure 4}
\end{figure}

Magnetoresistance of the sample heat treated in vacuum at $500^\circ C$ for
one hour is shown in Fig.\ \ref{Figure 4}. The unusual high field magnetoresistance cannot
be observed after the heat treatment but the resistance increases as
$[R(H)-R(0)]/R(0)\propto H^{1.5}$ in accordance with former results on Fe \cite{exponent}. 
The equality of $R_{\parallel}$ and $R_{\perp}$ has already made evident that 
the negative and
non-saturating high field magnetoresistance is not a thin film effect. The
disappearance of the high field anomaly after the heat treatment further
supports the idea that it arises from the granular nature of the interface
in the as-deposited sample. As it can be seen in Fig.\ \ref{Figure 4} the same heat
treatment removes the high field anomaly of the Cr/Fe/Cr trilayer, as well.
This is in accordance with the results on bulk supersaturated alloys \cite{O1}
where the recovery of the usual field dependence of the magnetoresistance
was attributed to the precipitation of large Fe clusters. 

On the basis of
the interface magnetoresistance observed in trilayers it is tempting to estimate the
magnitude of this effect in multilayers. If the mean free path of the
electrons is less than the layer thickness a parallel resistor network can
approximate the interface magnetoresistance of a multilayer. Supposing that
the interface conductivity is negligible in zero field but it gives a
significant contribution at the highest field applied, one obtains a value
around 7.7 kOhm and 5.4 kOhm for the high field resistance a single interface 
in the trilayer and the [$1.4nm$\ Fe\ +\ $2nm$\ Ag]$_{60}$  multilayer, respectively. 
The order of magnitude agreement in this simplified model and the 
good fit of the field dependence according to Eq.\ (\ref{eq:3}) suggest
that the magnetoresistance of Fe-Ag multilayers mainly arise from the
interface. 

In conclusion it is demonstrated for the first time that the
non-equilibrium structure of the interface in magnetic multilayers is the
source of a granular-type GMR behavior, which gives a contribution to the
magnetoresistance independently from the nature of coupling between the
layers. Separation of this interface magnetoresistance was possible by
studying a single magnetic layer sandwiched between nonmagnetic layers. 
To establish how the interface structure depends on the layer thickness and 
on the different parameters of the deposition technique need further studies. 
However, the granular interface magnetoresistance was shown to be dominant 
in policrystalline Fe/Ag multilayers. The observed magnitude of this 
effect in Cr/Fe/Cr trilayer indicates that the granular interface 
magnetoresistance can be non-negligible in Fe/Cr multilayers, as well.   

We thank M. Csontos and F. Z\'{a}mborszky for technical assistance and 
L. Bujdos\'{o} for sample preparation. This work 
was supported by Hungarian Research Funds OTKA T030753 and T026327.

\end{document}